\documentclass[
    ,final            
  ]
  {aipproc}

\layoutstyle{8x11single}

\newcommand{\ba}{\begin{eqnarray}}
\newcommand{\ea}{\end{eqnarray}}
\newcommand{\beqs}{\begin{eqnarray}}
\newcommand{\eeqs}{\end{eqnarray}}

\begin{document}

\title{The hadron structure and the description      of the elastic scattering in a wide region of $t$ and $s$}

\classification{{13.40.Gp}, 
      {14.20.Dh}, 
      {12.38.Lg} 
}
\keywords      {Elastic hadron scattering, hadron form factors, general parton distributions}

\author{O.V. Selyugin}{
  address={BLTP, JINR, Dubna, Russia}
}



\begin{abstract}
 On the basis of the new high energy general structure  (HEGS) model,
  which takes into account
  the  different moments of the  General Parton Distributions
  (GPDs) of the hadron,  the quantitative descriptions   of all
  existing experimental data from  $\sqrt{s} = 9.8$ GeV to 
  $  7 \ $ TeV, including
  the Coulomb range and large momentum transfers up to $-t=15$ GeV$^2$, are obtained with only
  3 free fitting high energy parameters.
  The real part of the hadronic amplitude is determined only
   through complex $s$ satisfying the dispersion relations.
The negligible contributions of the
  hard Pomeron and non-small contributions of the maximal Odderon is shown.
  Predictions for LHC energies are made.
\end{abstract}

\maketitle



\section{Introduction}
Now we present a model based on the assumption that the hadron interaction is sensitive to the generalized parton distributions, whose moments can be represented in the form of  two different distributions
 over momentum transfer:  charge and matter separately.
  Hence, this model uses the exact electromagnetic and  matter form-factors determined by one   function - generalized parton distributions (GPDs).
    So both form factors are independent of the fitting procedure
    of the differential elastic  cross sections.
    Note that the form of the GPDs is determined, on  the one hand, by   the deep-inelastic processes and, on the other hand, by the measure of the electromagnetic form factor from  the electron-nucleon elastic scattering.
 This picture is supported by a good description of the experimental data in the Coulomb-hadron interference region and a large momentum transfer at high energies by one amplitude with a few free parameters.
              The differential cross
  sections of the nucleon-nucleon elastic scattering  can be written as a sum of different
  helicity  amplitudes \cite{Rev-LHC}:
\begin{eqnarray}
  \frac{d\sigma}{dt} =
 \frac{2 \pi}{s^{2}} (|\Phi_{1}|^{2} +|\Phi_{2}|^{2} +|\Phi_{3}|^{2}
  +|\Phi_{4}|^{2}
  +4 | \Phi_{5}|^{2} ). \label{dsdt}
\end{eqnarray}
\renewcommand{\bottomfraction}{0.7}
  The total helicity amplitudes can be written as $\Phi_{i}(s,t) =
  F^{h}_{i}(s,t)+F^{\rm em}_{i}(s,t) e^{\varphi(s,t)} $\,, where
 $F^{h}_{i}(s,t) $ comes from the strong interactions,
 $F^{\rm em}_{i}(s,t) $ from the electromagnetic interactions and
 $\varphi(s,t) $
 is the interference phase factor between the electromagnetic and strong
 interactions \cite{Selphase}.
 For the hadron part the amplitude with spin-flip is neglected in this approximation, as usual at high energy.

\section{The High Energy General Structure (HEGS)  Model }
   We suppose that at high energies a hadron interaction in the non-perturbative regime
      is determined by the Reggeized-gluon exchange. The cross-even part of this amplitude can have two non-perturbative parts, possible standard pomeron - ($P_{2np}$) and cross-even part of the three-non-perturbative gluons ($P_{3np}$).
       The interaction of these two objects is proportional to two different form factors of the hadron.
      This is the main assumption of the model. The second important assumption is that we chose the slope of the second term
       four times smaller than the slope of the first term, by  analogy with the two pomeron cut.
      Both terms have the same intercept.
The electromagnetic form factors can be represented as the first  moments of GPDs
\ba
 F_{1}(t) = \int^{1}_{0}  dx  \ \sum_{u,d} {\cal{ H}}^{q} (x, t); \ \
 F_{2} (t) = \int^{1}_{0} dx \ \sum_{u,d} {\cal{E}}^{q} (x,  t),
\ea
  In \cite{GPD-ST-PR09}
 the $t$-dependence of  GPDs was chosen in the form
\ba
{\cal{H}}^{q} (x,t) \  = q(x)_{nf} \   \exp \left[  a_{+}  \
\frac{(1-x)^2}{x^{m} } \ t \right]\;;  \ \ \ \ \
{\cal{E}}^{q} (x,t) \  = q(x)_{sf} \   \exp \left[  a_{-}  \
\frac{(1-x)^2}{x^{m} } \ t \right]\;.
\ea
The function $q(x)$ was taken at the scale $\mu^2=1$ 
 and was based on the MRST2002 global fit \cite{MRST02}.
 For $\xi=0 $ one has
\ba
\int^{1}_{0} \ dx \ x \sum_{u,d}[{\cal{H}}(x,t) \pm {\cal{E}}(x,t)] = A_{}(t) \pm B_{}(t)\;.
\ea
The integration of the second momentum of GPDs over $x$ gave  the momentum-transfer representation
  of the form factor.  
  It was  approximated  by the dipole form \cite{HEGS0-JEP12}
  $ \ A(t)=L^{4}_{2}/(L^{2}_{2}-t)^2 $ .   
      Hence, the Born term of the elastic hadron amplitude can be written as
  \begin{eqnarray}
 F_{h}^{Born}(s,t) \ =  h_1 \ G^{2}(t) \ F_{a}(s,t) \ (1+r_1/\hat{s}^{0.5})  
    \    +  h_{2} \  A^{2}(t) \ F_{b}(s,t) \ (1+r_2/\hat{s}^{0.5})\;,
\end{eqnarray}
  where $F_{a}(s,t)$ and $F_{b}(s,t)$  has the standard Regge form 
  \begin{eqnarray}
 F_{a}(s,t) \ = \hat{s}^{\epsilon_1} \ e^{B(s) \ t}; \ \ \
 F_{b}(s,t) \ = \hat{s}^{\epsilon_1} \ e^{B(s)/4 \ t}\;.
\end{eqnarray}
  The slope of the scattering amplitude has the standard logarithmic dependence on the energy.
 $   B(s) = \alpha_{1} \ \ln(\hat{s})$,
  with $\alpha_{1}=0.24$ GeV$^{-2}$ and   $  \hat{s}=s \ e^{-i \pi/2}/s_{0} ;  \ \ \ s_{0}=1 \ {\rm GeV^2} $.
 The final elastic  hadron scattering amplitude is obtained after unitarization of the  Born term.
    So, first, we have to calculate the eikonal phase
   \begin{eqnarray}
 \chi(s,b) \   =  \frac{1}{2 \pi}
   \ \int \ d^2 q \ e^{i \vec{b} \cdot \vec{q} } \  F^{\rm Born}_{h}\left(s,q^2\right)\,,
 \label{tot02}
 \end{eqnarray}
  and then obtain the final hadron scattering amplitude
    \begin{eqnarray}
 F_{h}(s,t) = i s
    \ \int \ b \ J_{0}(b q)  \ \Gamma(s,b)   \ d b\, ;  \ \ \  \ \ \
  \Gamma(s,b)  = 1- \exp[- \chi(s,b)] .
 \label{overlap}
\end{eqnarray}
   All these calculations are carried out by the FORTRAN program. The model has only three high energy fitting parameters
 and $2$ low energy parameters, which reflect some small contribution
 coming from the different low energy terms.
  We take all existing
      experimental data in the energy range $52.8 \leq \sqrt{s} \leq 1960 \ $GeV
      and the region of the momentum transfer $0.0008 \leq \ -t \ \leq 9.75 \ $GeV$^2$
      of the elastic differential cross sections of proton-proton and proton-antiproton
      data \cite{data-Sp}.     
As a result, one obtains $\sum \chi^2_i /N \simeq 1.8 $ where $N=975 $ is
the number of experimental points.
Note that the parameters of the model are energy-independent.
The energy dependence of the scattering amplitude is determined
only by the single intercept and the logarithmic dependence on $s$ of the slope.
\section{Extension of the HEGS model}
   Further development of the model requires  careful analysis of the $t$ form of GPDs and
   a properly chosen form of PDFs. In \cite{GPDs-PRD14}, the analysis of more than 20 different
   PDFs was made. We slightly complicated the form of GPDs
\ba
{\cal{H}}^{u} (x,t) \  = q(x)_{nf} \   \exp \left[ 2 a_{+}  \ \frac{(1-x)^{2+\epsilon_{1}}}{x^{m}}  \ t \right]\;; \ \  \ \ \ \
{\cal{H_d}}^{d} (x,t) \  = q(x)_{nf} \   \exp \left[ 2 a_{+}  \ (\frac{(1-x)^{2+\epsilon_{2}} + d x (1-x)}{x^{m}} ) \ t \right]\;,
\ea
\ba
{\cal{E}}^{u} (x,t) \  = q(x)_{nf} \   \exp \left[ 2 a_{+}  \ \frac{(1-x)^{2+\epsilon_{1}}}{x^{m}}  \ t \right]\;; \ \  \ \ \ \
{\cal{E_d}}^{d} (x,t) \  = q(x)_{nf} \   \exp \left[ 2 a_{+}  \ (\frac{(1-x)^{2+\epsilon_{2}} + d x (1-x)}{x^{m}} ) \ t \right]\;.
\ea
      On the basis of this analysis we calculated the nucleon electromagnetic form factor and
      the matter form factors.
      The fit of these calculations gave
      the parameters of the electromagnetic $G(t)$ and matter form factors $A(t)$.

      Then, as we intend to describe sufficiently low energies, the possible odderon contributions
      were taken into account.
  \begin{eqnarray}
 F_{odd}^{Born}(s,t) \ =  \pm i h_{odd} \ A^{2}(t) \ \frac{t}{1-r_{0}^{2} t } \ F_{b}(s,t).
\end{eqnarray}
   As we supposed in the previous variant of the HEGS model that the $F_{b}(s,t) A(t)^2$ correspond
   to the cross-even part of the three gluon exchange, our odderon contribution is also connected
   with the  matter form factor $A(t)$. Our ansatz for the odderon slightly differs from the cross-even part by some kinematic factor.
 The form of the odderon works in all $t$ 
  and  has the same behavior at non-small $t$ as the cross-even part, of course, with changing sign for proton-proton and proton-antiproton reactions.

  We added a small contribution of the asymptotic energy independent   part of the spin flip amplitude in the form similar to the  proposed    in \cite{Kuraev-SF}.
 $ F_{sf}(s,t) \ =  h_{sf} q^3 G^{2}(t)$ .
  It has only one additional free parameter.

 In our new description of the experimental data of the proton-proton and proton-antiproton
 we took the energy region from $\sqrt{s} > 9.8 $ GeV.
 It gave us many new high precision  experimental data especially at small $t$ 
  in the  CNI 
   region.
   Hence, we need to more careful by examine the slope of the scattering amplitude.
   Taking into account the work \cite{Jenk}, we take some small nonlinear form
  of the slope of the scattering amplitude with  the standard logarithmic dependence on the energy.
 $   B(s) = (\alpha_{1} + q k_{0}e^{k_{0} t} ) \ \ln(\hat{s}) $,
  with $\alpha_{1}=0.24$ GeV$^{-2}$
and
$\Delta=0.11; s_{0}=4 m_{p}^{2} $ - fixed. 
     Our complete fit of  3088 experimental data in the energy range
      $9.8 \leq \sqrt{s} \leq 7000 \ $ GeV
      and the region  of the momentum transfer
      $0.00045 \leq \ -t \ \leq 14.75 \ $GeV$^2$ gave
  $ \ \ \sum_{i=1}^{N} \chi_{i}^{2}/N=1.33$
with the parameters
$h_1=0.82; \ \ h_2=0.31; \ \  h_{odd}=0.15; \ \
 k_0=0.17; \ \  r_{0}^{2}=3.9 \ \ \ $ and the low energy parameters
 $ \ h_{sf} = 0.05; \ \ R_1=51.$ and $R_2 = 4.4$.

The model reproduces $d\sigma/dt$ at very small and large $t$ and provides a qualitative description of the dip region
at $-t \approx 1.4 $~GeV$^2 $, for $\sqrt{s}=53 $~GeV$^2 $ and for $\sqrt{s}=62.1 $~GeV$^2 $.
  Noteworthy is a good description of the CNI region in a very wide energy region,
   approximately three orders, with the same slope of the scattering amplitude.
 The differential cross sections of the elastic scattering  $pp$ and $p\bar{p}$
  at small $t$ and different $s$ are presented in
  Fig.1.  for  $\sqrt{s}= 9.8 $ GeV and $7$ TeV $(pp)$ and $\sqrt{s}= 11. $ GeV  $(p\bar{p})$.
   The model quantitatively reproduces the differential cross sections in the whole examined energy region
 in spite of  the size of the slope changing essentially in this region (due to the standard Regge behavior $\ln(\hat{s})$)  and of the real part of the scattering amplitude having the different behavior for
   $pp$ and $p\bar{p}$.

In Fig.~2, the description of the differential cross sections of the elastic scattering  $pp$ and
   $p\bar{p}$  at large $t$ and different $s$ is presented.
   It should be noted that the calculations of our  integrals with complex oscillation functions at large $t$ is a difficult task    and requires high precision of the calculations. In any case,
   we obtain a quantitatively good description of the differential cross sections
   at  large $t$.
   In this region of $t$ the contribution of the spin-flip amplitude is visible.
   We take into account only the asymptotic part of this amplitude with the simplest and  energy independent
    forms. Although it has a small size, its constant is determined sufficiently well
     $h_{sf} =0.05 \pm 0.002$.


\begin{figure}
  \includegraphics[height=.2\textheight]{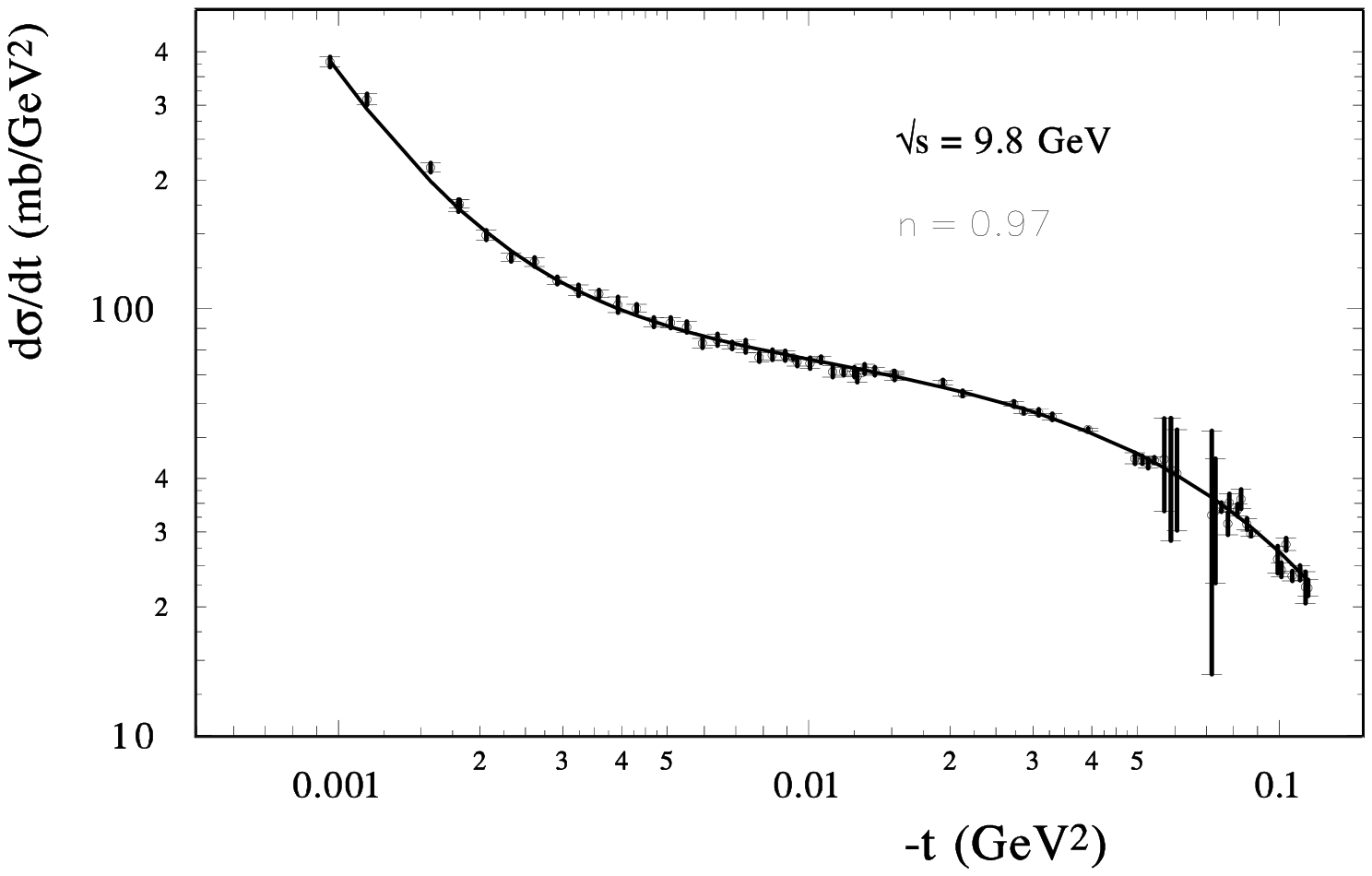}
      \includegraphics[height=.2\textheight]{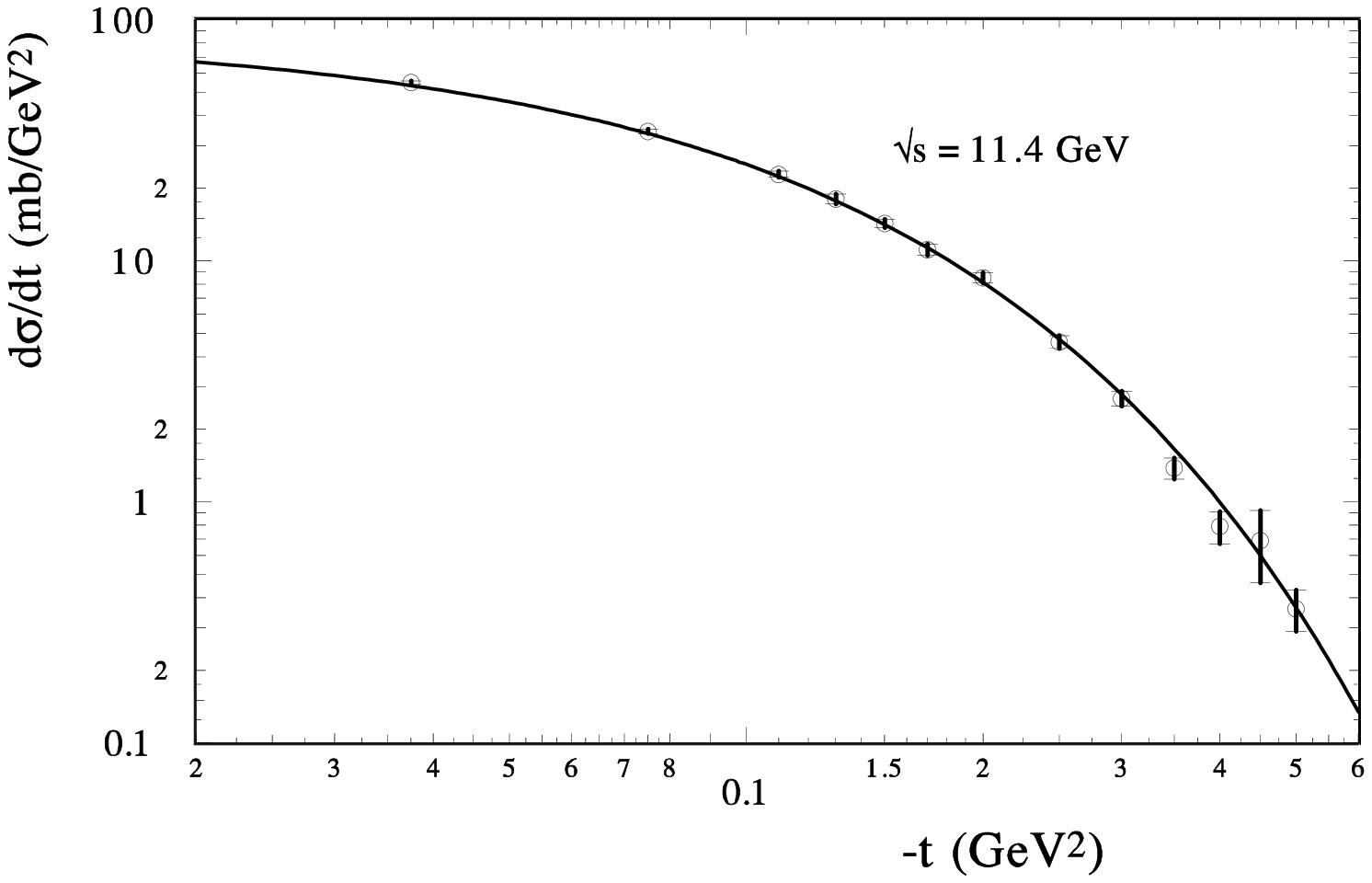}
    \includegraphics[height=.2\textheight]{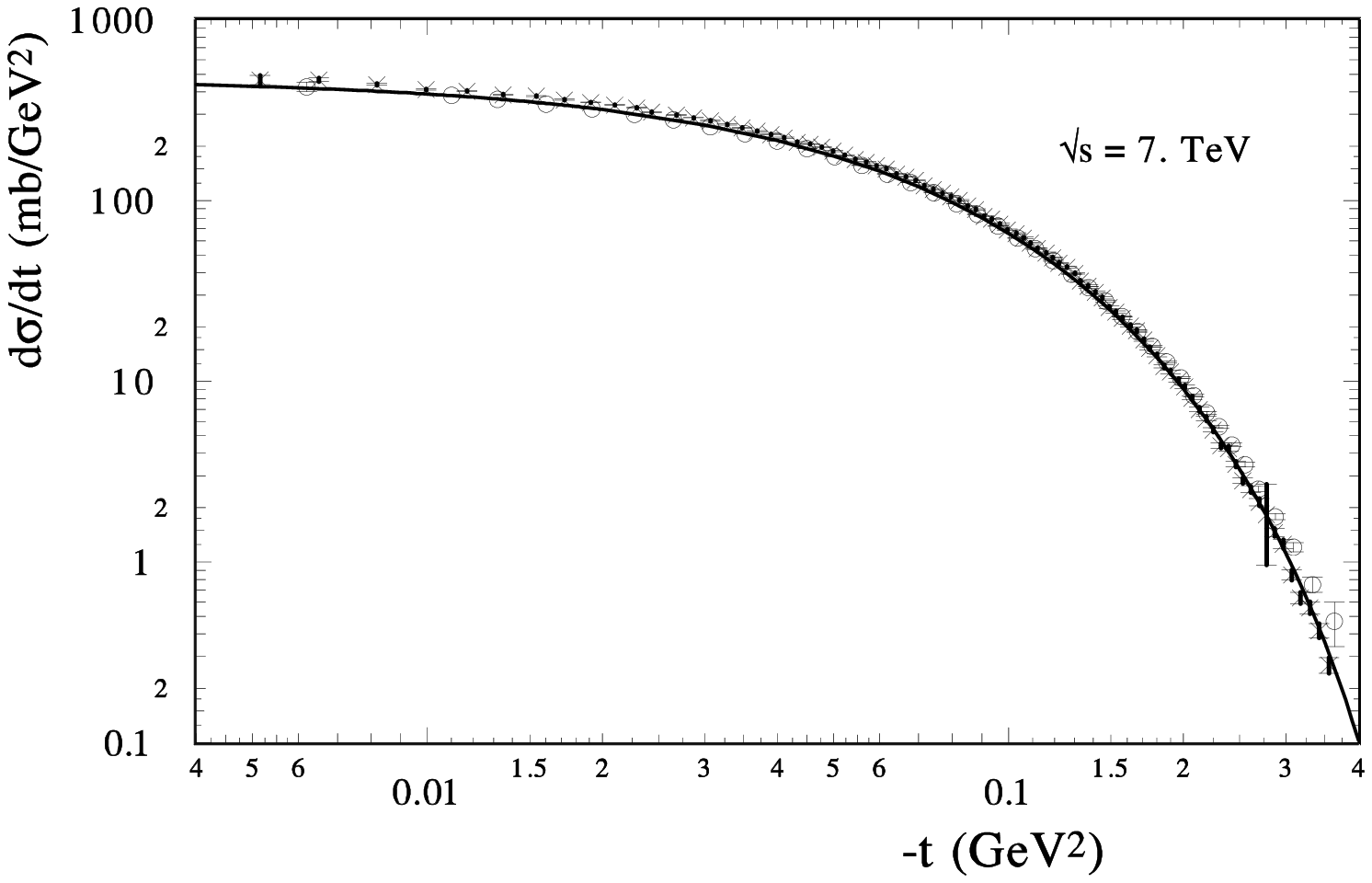}
  \caption{
  $d\sigma/dt$ a) for $pp$ (left panel) at $\sqrt{s}=9.8$ GeV
    and  b) $p\bar{p}$ (central panel) at $\sqrt{s}=11.3$ GeV  and c)
      $pp$ (right panel) at $\sqrt{s}=7$ TeV [cross - TOTEM data [10], circles - ATLAS data [11].
   }
\end{figure}

\begin{figure}
  \includegraphics[height=.2\textheight]{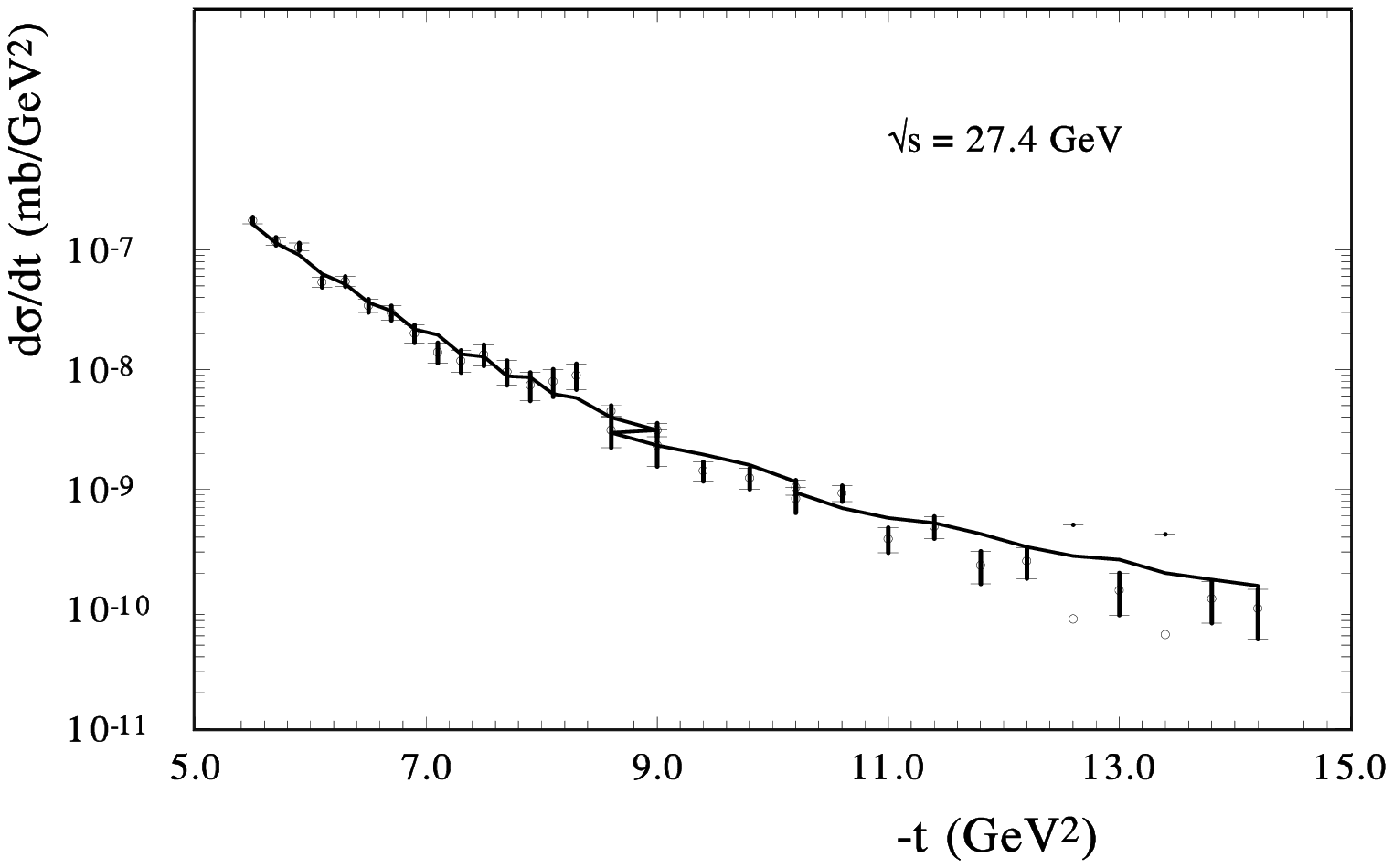}
      \includegraphics[height=.2\textheight]{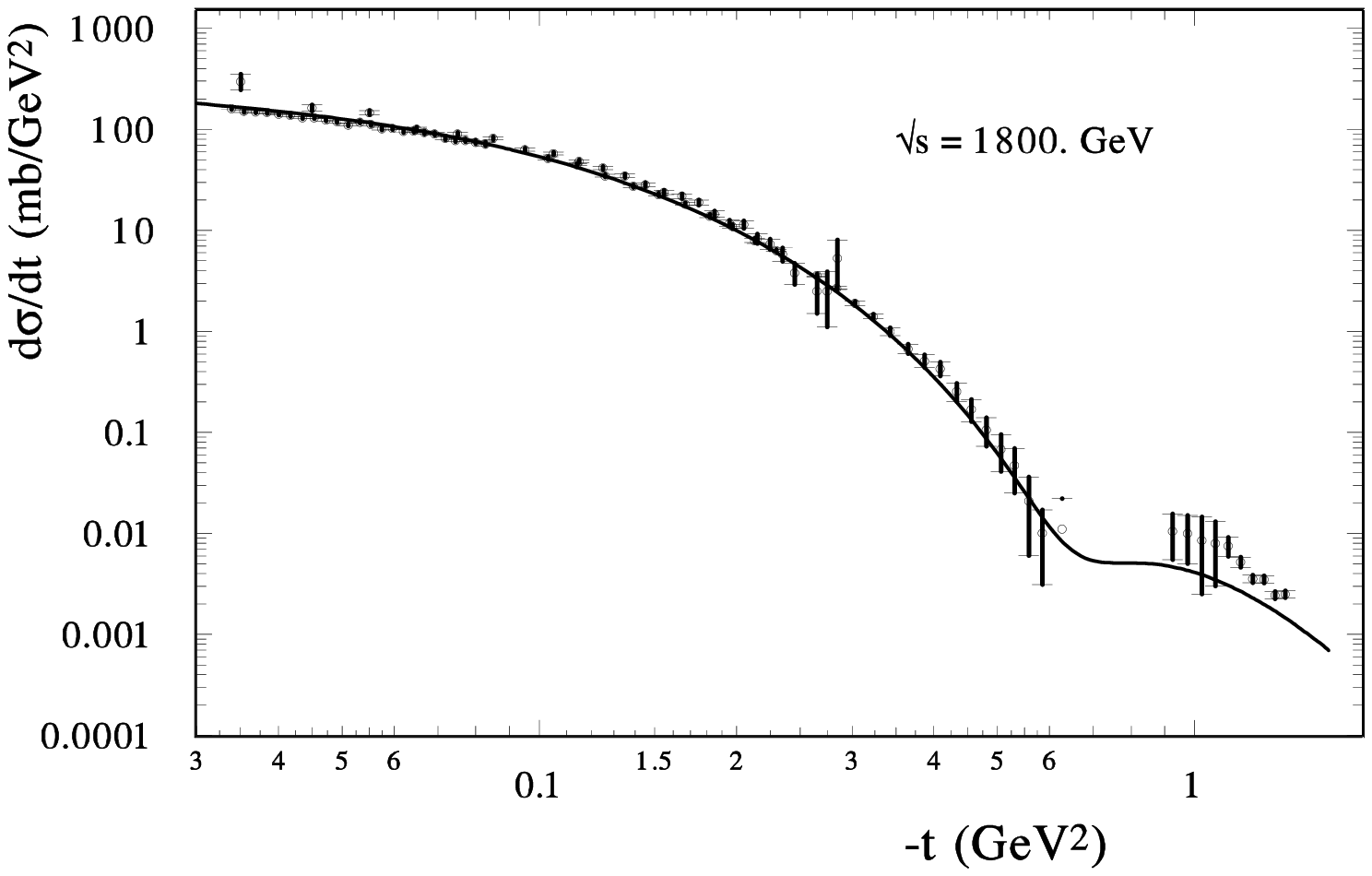}
    \includegraphics[height=.2\textheight]{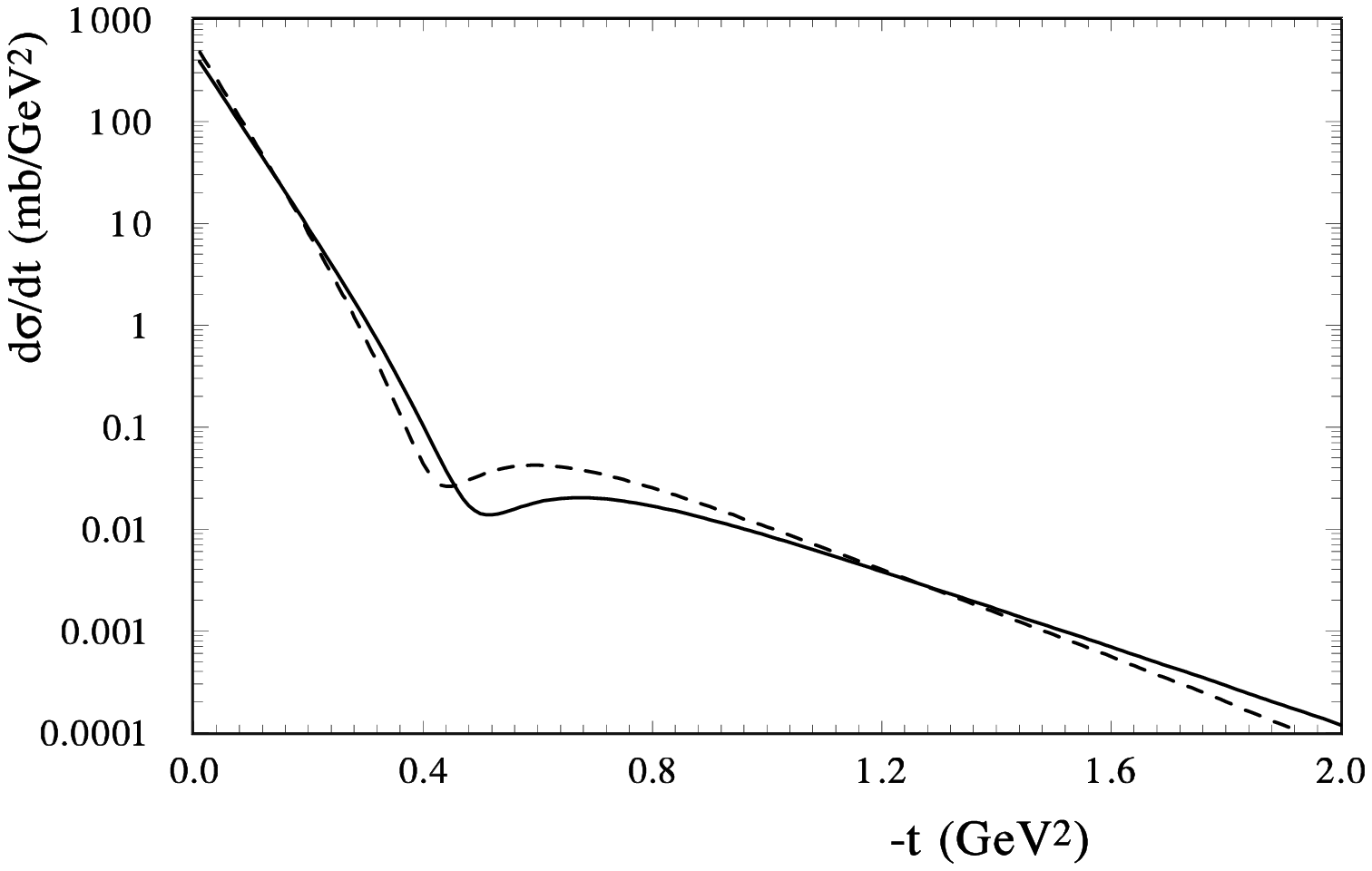}
  \caption{$d\sigma/dt$ at large $t$ a) for $pp$ (left panel) at $\sqrt{s}=27.43$ GeV
    and  $p\bar{p}$ (central panel) at $\sqrt{s}=1.8$ TeV
    and the model predictions at $\sqrt{s}=7$ and $14$ TeV. }
\end{figure}

\section{Conclusions}

  We presented a new model of the hadron-hadron interaction at high energies.
  The model is very simple from the viewpoint of the number of parameters and functions.
  There are no artificial functions or cuts which bound the separate
  parts of the amplitude by some region of momentum transfer or energy.
  One of the most remarkable properties is that the real part of the
  hadron scattering amplitude is determined only by complex energy $\hat{s}$ that satisfies the crossing-symmetries.
 The new HEGS
  model  gives the quantitatively  description of the elastic nucleon scattering at high energy  with only six fitting high energy parameters.
 Our model of GPDs leads to the well description of the proton and neutron  electromagnetic form factors and its elastic scattering simultaneously.
   The successful description  of the existing experimental data by the model show that
   the elastic scattering determined by the generalized structure of the hadron.
The model leads to a good coincidence of the model calculations with the preliminary data at 7 TeV.
   We find that the  standard eikonal approximation \cite{Unit-PRD}   works perfectly well from
  $\sqrt{s}=9$  GeV up to  7 TeV.
The extending variant of the model shows the contribution of the "maximal" odderon with specific
   kinematic properties
  and does not show a visible contribution of the hard pomeron \cite{NP-HP}.

The  slope of the differential cross sections at small $t$ has a small
      peculiarity and has the same properties in the whole energy
      region examined. Such a uniform picture for the slope gives a possibility to further research
      the small peculiarity of the different cross sections like some possible oscillations \cite{Rev-LHC,AKM-PL,LRPot-Dif10}.
   Note that  we do not see contributions of the second Reggeon with a large slope and intercept above 0.5.

      Including  the data of the TOTEM Collaboration in our fit increases
       $\sigma_{tot}$  at $\sqrt{s}=7$ TeV  from  $95$ mb \cite{HEGS0-JEP12}    to $97$ mb now (see also \cite{TOTEMrho-NPhis14}).


\begin{theacknowledgments}
O.S. would like to thank the organizers R. Fiore and A. Papa for the invitation
and support of his participation in the conference.
\end{theacknowledgments}



\bibliographystyle{aipproc}   


%


\end{document}